\documentstyle[12pt]{article} \long\def\del#1{ }
\let\3=\ss \catcode`\"=\active \let"=\"
\let\ssk=\smallskip  
  
   \def\ve{\vfil\eject}

\let\a=\alpha \let\b=\beta   \let\e=\varepsilon
    
\let\l=\lambda \let\m=\mu

\def\0{\over }    \def\1{\vec }   \def\2{{1\over2}} \def\3{{\ss}}
\def\4{{1\over4}} \def\5{\bar }   \def\6{\partial } \def\7#1{{#1}\llap{/}}
\def\8#1{{\textstyle{#1}}}        \def\9#1{{\bf {#1}}}
\def\_#1{$\underline{\hbox{#1}}$} \def\~#1{$\overline{\hbox{#1}}$}

\def\<{\langle } \def\>{\rangle }  
 
\def \({\left( } \def \){\right) }

 \let\eq=\equiv    
      \let\and=\wedge
     
\def\|#1{{}_{\bigg|_{#1}}}

\def\mao#1{\mathop{\rm {#1}}\nolimits}  \def\tr{\mao{tr}}

\def\pmbf#1{\setbox0=\hbox{${#1}$}   \kern-.025em\copy0\kern-\wd0
      \kern.05em\copy0\kern-\wd0     \kern-.025em\raise.0433em\box0 }
   
   \def\ch{{\cal H}}
  
\def\inbar{\vrule height1.5ex width.4pt depth0pt}
\def\IN{\relax{\rm I\kern-.18em N}}    \def\ZZ{\relax{\sf Z\kern-.4em Z}}
\def\IP{\relax{\rm I\kern-.18em P}}
\def\IQ{\relax\,\hbox{$\inbar\kern-.3em{\rm Q}$}}
\def\IR{\relax{\rm I\kern-.18em R}}
\def\IC{\relax\,\hbox{$\inbar\kern-.3em{\rm C}$}}

\def\beq{\begin{equation}} \def\eeq{\end{equation}} \def\eql#1{\label{#1}\eeq}
\def\bea{\begin{eqnarray}} \def\eea{\end{eqnarray}} 
\def\fnote#1#2{\begingroup\def\thefootnote{#1}\footnote{#2}
	     \addtocounter{footnote}{-1}\endgroup}

\def\plb#1 #2 {Phys. Lett. {\bf B#1} #2 }
\def\phr#1 #2 {Phys. Rep. {\bf  #1} #2 } 
\def\npb#1 #2 {Nucl. Phys. {\bf B#1} #2 }
\def\aph#1 #2 {Ann. Phys. {\bf #1} #2 }  
\def\jmp#1 #2 {J. Math. Phys. {\bf #1} #2 }
\def\prd#1 #2 {Phys. Rev. {\bf D#1} #2 }
\def\prl#1 #2 {Phys. Rev. Lett. {\bf #1} #2 }
\def\rmp#1 #2 {Rev. Mod. Phys.  {\bf #1} #2 }
\def\zpc#1 #2 {Z. Phys. {\bf #1C} #2 }
\def\cmp#1 #2 {Comm. Math. Phys. {\bf #1} #2 }
\def\mpl#1 #2 {Mod. Phys. Lett. {\bf A#1} #2 }
\def\ijmp#1 #2 {Int. J. Mod. Phys. {\bf A#1} #2 }

\def\ch{\hat c} \let\chat=\ch   
\def\tbf#1:{{\noindent\bf #1:}}  

\def\figuresonly{\pagestyle{empty}\figa\figb\figc\figd\fige\end{document}}
\def\old#1\endold{{ }}
\newcount{\npre}  \npre=1  \def\negansw{n } \def\figansw{f } \def\oldansw{o }
\def\ifpre{\ifnum\npre>0 } \def\ifsub{\ifnum\npre=0 }
\def\askversion{\message{
    Preprint version (y)  <<or: Figures appended (n)>>  (y/n)? }
    \read-1 to\answ  \ifx\answ\negansw \npre=0 \else \npre=1 \fi
    \ifx\answ\figansw { } \else \def\figuresonly{ }          \fi
    \ifx\answ\oldansw \def\old##1\endold{{\small ##1} } \fi  \figuresonly}
\def\putfig#1{{\ifpre\begin{figure}{#1}\end{figure}\fi}}

\def\bpic{\begin{picture}} \def\epic{\end{picture}} \thicklines
\def\pt#1){\put#1){\circle*{3}}}
\def\lab#1)#2#3{\put#1){\makebox(0,0)[#2]{\small #3}}}
\def\putlin#1,#2,#3,#4,#5){\put#1,#2){\line(#3,#4){#5}}} 
\def\putvec#1,#2,#3,#4,#5){\put#1,#2){\vector(#3,#4){#5}}}
\def\postri#1{\multiput(0.125,0.25)(0.25,0.5){#1}{\circle*{1}}} 
\def\posdia#1{\multiput(0.25,0.25)(0.5,0.5){#1}{\circle*{1}}}   
\def\negtri#1{\multiput(0.125,-.25)(0.25,-.5){#1}{\circle*{1}}} 
\def\negdia#1{\multiput(0.25,-.25)(0.5,-0.5){#1}{\circle*{1}}}  
\newsavebox{\pointer}                                                
\savebox{\pointer}(60,80){\bpic(60,80)(-30,-20)
   \pt(0,0) \lab(0,-10)b2  \pt(0,40) \lab(0,50)t1
   \put(0,40){\vector(0,-1){20}} \put(0,20){\line(0,-1){20}} \epic}
\newsavebox{\Loop}
\savebox{\Loop}(100,80){\bpic(100,80)(-30,-20)
   \pt(20,20) \lab(10,20)l3 \pt(80,20) \lab(90,20)r4
   \put(50,20){\oval(60,40)}
   \put(52,0){\vector(1,0){0}} \put(48,40){\vector(-1,0){0}} \epic}
\def\figa{\vbox{
\begin{center}                                               
\bpic(160,80) \put(0,0){\usebox{\pointer}} \put(40,0){\usebox{\Loop}} \epic
\end{center}
Fig. 1: Graphic representation of $X_1^{\a_1}X_2+X_2^{\a_2}+
      X_3^{\a_3}X_4+X_4^{\a_4}X_3$.}      }

\newsavebox{\vier}                                                   
\savebox{\vier}(80,80){\bpic(80,80)(-30,-20)
   \pt(0,0)\lab(0,-10)b2  \pt(-20,40)\lab(-25,50)t1  \pt(20,40)\lab(25,50)t3
   \put(-20,40){\vector(1,-2){10}} \put(-10,20){\line(1,-2){10}}
   \put(20,40){\vector(-1,-2){10}} \put(10,20){\line(-1,-2){10}}
   \pt(50,0) \lab(50,-10)b4  \epic}
\def\figb{\vbox{
\begin{center}                                               
\bpic(80,80) \put(0,0){\usebox{\vier}}              \epic
\end{center}
Fig. 2: Graphic representation of the degenerate polynomial
      \hbox{$X_1^{\a_1}X_2+X_2^{\a_2}+X_3^{\a_3}X_2+X_4^{\a_4}$}.}   }

\newsavebox{\wigg}                                                 
\savebox{\wigg}(80,80){\bpic(80,80)(-30,-20) \put(-30,-20){\usebox{\vier}}
       \multiput(-20,40)(6.6,0){6}{\bpic(6.6,0)\put(0,0){\line(1,0){2}}
	 \put(5.6,0){\line(1,0){2}}\epic}                  \epic}
\newsavebox{\poin}
\savebox{\poin}(80,80){\bpic(80,80)(-30,-20) \put(-30,-20){\usebox{\wigg}}
   \put(25,40){\oval(50,35)[t]}   
   \put(50,40){\vector(0,-1){20}} \put(50,20){\line(0,-1){20}}    \epic}
\def\figc{\vbox{
\begin{center}                                               
\bpic(170,80) \put(-40,0){\usebox{\wigg}} \put(140,0){\usebox{\poin}} \epic
\end{center}
Fig. 3: Dressing of the graph in Fig. 2:
      (a) by the link $X_1^{p}X_3^{q}$
      and (b) by the pointer $X_1^{p}X_3^{q}X_4$.   }}

\newsavebox{\pdia} \savebox{\pdia}(3,3){\bpic(0,0)\posdia6\epic}
\newsavebox{\ndia} \savebox{\ndia}(3,-3){\bpic(0,0)\negdia{6}\epic}
\def\figd{\vbox{
\begin{center}                                               
\bpic(160,100)(-20,-20)
   \pt(0,0)\lab(-10,-10){bl}2 \pt(0,40)\lab(-10,40)l1 \pt(40,0)\lab(40,-10)b3
   \put(0,0){\line(0,1){20}}  \put(40,0){\vector(-1,0){20}}
   \put(0,0){\line(1,0){20}}  \put(0,40){\vector(0,-1){20}}
   \multiput(2,38)(6,-6){6}{\usebox{\ndia}}
\pt(120,0)\lab(130,-10){br}5 \pt(120,40)\lab(130,40)r6 \pt(80,0)\lab(80,-10)b4
   \put(120,0){\line(0,1){20}}  \put(80,0){\vector(1,0){20}}
   \put(100,0){\line(1,0){20}}  \put(120,40){\vector(0,-1){20}}
   \multiput(82,2)(6,6){6}{\usebox{\pdia}}
\pt(60,60)\lab(60,70)t7
   \put(20,20){\vector(1,1){20}}  \put(40,40){\line(1,1){20}}
   \put(100,20){\vector(-1,1){20}} \put(80,40){\line(-1,1){20}}
\multiput(20,20.4)(8,0){10}{\bpic(0,0)\put(2,0){\line(1,0){4}}\epic}
\multiput(20,19.4)(8,0){10}{\bpic(0,0)\put(2,0){\line(1,0){4}}\epic}
\epic
\end{center}
Fig. 4: The pointers $(13)\to7$ and $(46)\to7$ require the link ((13)(46))
      drawn in boldface. }  }

\newsavebox{\ptri} \savebox{\ptri}(2,4) {\bpic(0,0)\postri8\epic}
\newsavebox{\ntri} \savebox{\ntri}(2,-4){\bpic(0,0)\negtri8\epic}
\def\fige{\vbox{
\begin{center}                                               
\bpic(160,100)(-80,-40)         \pt(0,20)\lab(0,10)b1
   \pt(-30,0)\lab(-40,-10){bl}2 \putvec(-30,0,3,2,15) \putlin(-15,10,3,2,15)
   \pt(30,0)\lab(40,-10){br}3   \putvec(30,0,-3,2,15) \putlin(15,10,-3,2,15)
   \pt(0,60)\lab(0,70)t4    \putvec(0,60,0,-1,20) \putlin(0,40,0,-1,20)
   \pt(0,-20)\lab(0,-30)b5  \putvec(0,0,0,-1,10)  \putlin(0,-10,0,-1,10)
   \put(-15,5){\oval(70,50)[l]} \putvec(-50,5,0,-1,0)
   \put(15,5){\oval(70,50)[r]}  \putvec(50,5,0,-1,0) \putlin(-15,-20,1,0,30)
\multiput(-31,-2)(4,8){8}{\usebox{\ptri}}
\multiput(-1,62)(4,-8){8}{\usebox{\ntri}}
\multiput(-30,0)(8,0){8}{\line(1,0){4}}
\multiput(-15.5,31)(4,-8){4}{\usebox{\ntri}}
\multiput(-14.7,31)(4,-8){4}{\usebox{\ntri}}  
\multiput(-.5,-1)(4,8){4}{\usebox{\ptri}}
\multiput(-1.3,-1)(4,8){4}{\usebox{\ptri}}    
\multiput(-14,30)(8,0){4}{\line(1,0){4}}
\multiput(-14,29.2)(8,0){4}{\line(1,0){4}}    
\epic
\end{center}
Fig. 5: The monomial $X_2^pX_3^qX_4^r$ provides all three links ((23)(34)),
      ((34)(42)) and ((42)(23)) required by the pointers $(23)\to5$,
      $(34)\to5$, and $(42)\to5$. }    }
\begin{document} \askversion                    {\hfill CERN-TH.6373/92}
\vskip -3mm                                          {\hfill TUW--92--1}
\vskip 15mm
\begin{center}{\large ON THE CLASSIFICATION OF QUASIHOMOGENEOUS FUNCTIONS}
\vskip 12mm
       Maximilian KREUZER\fnote{\star}{e-mail: kreuzer@cernvm.cern.ch}
\vskip 5mm
	  Theory Division, CERN\\
       CH--1211 Geneva 23, SWITZERLAND
\vskip 7mm               and
\vskip 7mm        Harald SKARKE\fnote{\#}{e-mail: hskarke@email.tuwien.ac.at}
\vskip 5mm
      Institut f"ur Theoretische Physik, Technische Universit"at Wien\\
	  Wiedner Hauptstrasse 8--10, A--1040 Wien, AUSTRIA

\vfil                        {\bf ABSTRACT}                \end{center} \ssk

   We give a criterion for the existence of a non-degenerate quasihomogeneous
polynomial in a configuration, i.e. in the space of polynomials
with a fixed set of weights, and clarify
the relation of this criterion to the necessary condition derived from the
formula for the Poincar\'e polynomial. We further prove finiteness of the
number of configurations for a given value of the singularity index. For
the value 3 of this index, which is of particular interest in string theory,
a constructive version of this proof implies an algorithm for the calculation
of all non-degenerate configurations.

\vfil \noindent
CERN-TH.6373/92\\
TUW--92--1\\
January 1992
\thispagestyle{empty} \eject \setcounter{page}{1} \pagestyle{plain}
\ifsub \baselineskip=20pt \else \baselineskip=18pt \fi

\section{Introduction}

Recently a particular class of singularities \cite{a,agv}, namely
singularities of holomorphic
quasihomogeneous functions, have been found useful for the
classification of superconformal field theories (SCFT)
with particular significance for the case of $N=2$ superconformal symmetry
\cite{m,lvw} due to a non-renormalisation theorem.
The requirement of conformal invariance implies
quasihomogeneity of degree 1 for the superpotential
\beq  W(\l^{n_i}\Phi_i)=\l^d W(\Phi_i)                               \eql{qh}
in the effective Lagrangian description, with the
scaling dimensions of the chiral superfields $\Phi_i$ translating into the
weights $q_i=n_i/d$ of the variables $X_i$ of a holomorphic function $W(X_i)$.
The local algebra of this function, i.e. the quotient of the polynomial ring
by the ideal generated by the gradients $\6_jW(X_i)$, is isomorphic to the
operator product algebra of chiral primary states \cite{lvw}. In order that
this algebra be finite we will concentrate on isolated singularities, i.e.
require that the origin be the only solution to the equation $dW=0$.
We will consider only quasihomogeneous holomorphic functions with positive
weights $q_i>0$, which are automatically polynomials.

In the mathematical literature isolated singularities have been classified
up to 3 variables and for low values of
the modality or of the dimension of the local algebra \cite{agv}.
A general tool for the investigation of degeneracy is the Poincar\'e
polynomial $P(t)=\tr t^{dJ}$ \cite{agv,bou}, where the trace extends
over a basis of the
local algebra and $J=\sum_iq_iX_i\6_i$ gives the weight of a basis monomial,
i.e. the coefficients $\m_a$ of $t^{dq_a}$ in this polynomial are the
multiplicities of the weights $q_a$.
For any non-degenerate quasihomogeneous function this
polynomial can be calculated from
\beq P(t)=\prod_i{1-t^{d-n_i}\01-t^{n_i}}.                            \eql{pp}
Thus a necessary condition for non-degeneracy is that this expression is a
polynomial with non-negative coefficients. Note that the Poincar\'e polynomial
only depends on the set of weights and not on the particular form of $W$.
We call the set of all polynomials which are quasihomogenous with respect
to the weights $q_i=n_i/d$ a configuration
\beq             \IC_{(n_1,\dots ,n_I )}[d].                        \eql{aff}
A configuration is said to be degenerate if it has only degenerate members.
A final guidance of our interest comes
from the relation between the singularity index
\beq               \hat c=\sum_i(1-2q_i)                      \eql{lgp}
and the central charge c in the Virasoro subalgebra of the corresponding
N=2 SCFT, which can be shown to be $c=3\hat c$ \cite{lvw}.

In this note we will supply tools for the classification of non-degenerate
configurations with a given singularity index.
In section 2 we start with some definitions and examples and then give
a criterion for non-degeneracy of a configuration. We also show how this
criterion is related to the necessary condition that the r.h.s. of
(\ref{pp}) is a
polynomial (PP-condition).
In section 3 we prove that for any given value of the singularity index
there is only a finite number of non-degenerate configurations.
Then we consider the cases of low integer index $\ch\le3$ more explicitly.
In the conclusion we summarize our results and briefly comment on their
implications for the classification of SCFTs.

\section{Non-degeneracy criterion}
A necessary condition for a quasihomogeneous polynomial
to be non-degenerate is that every variable $X$ occurs either in the
form $X^\a$ or $X^\a Y$. We will use a graphic representation
in which every variable is represented by a dot, and a term of the form
$X^\a Y$ is indicated by an arrow from $X$ to $Y$. We will sometimes say
`$X$ points at $Y$'.
If no arrow originates from a variable $X$, then there is a term of the
form $X^\a$ in the polynomial. For example, a polynomial
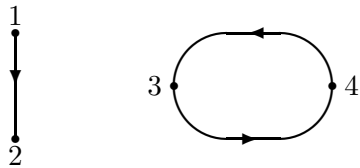
of the form $X_1^{\a_1}X_2+X_2^{\a_2}+X_3^{\a_3}X_4+X_4^{\a_4}X_3$
is shown in Fig.~1.
\putfig\figa

A possible danger for the non-degeneracy of the polynomial arises
when two or more arrows end at the same point, such as in the polynomial
$X_1^{\a_1}X_2+X_2^{\a_2}+X_3^{\a_3}X_2+X_4^{\a_4}$, depicted in Fig.~2.
\putfig\figb
The reason for the problem is that when we calculate $dW/dX_i$ and
set $X_2$ to $0$ we are left with only one equation for the two
variables $X_1$ and $X_3$. Therefore we have to introduce extra
terms in $W$ that yield terms in $X_1$ and $X_3$ in $dW$ even when all
other variables are set to zero, as in
$X_1^{\a_1}X_2+X_2^{\a_2}+X_3^{\a_3}X_2+X_4^{\a_4}+\e X_1^pX_3^q$ or in
$X_1^{\a_1}X_2+X_2^{\a_2}+X_3^{\a_3}X_2+X_4^{\a_4}+\e X_1^pX_3^qX_4$.
The graphic representation of these polynomials
is shown in Fig.~3.
\putfig\figc
We use dashed lines and pointers originating from dashed lines to
indicate such
extra terms. We shall call the graph without these additional lines
``skeleton graph''.
It should be noted that
all weights can be calculated only with the knowledge of the skeleton.
As we have already mentioned,
a necessary condition for the existence of a non-degenerate quasihomogeneous
function is that the r.h.s of (\ref{pp})
is actually a polynomial. For configurations
with up to three variables, but not in general, this condition is also
sufficient and identical to the condition for the existence of the
terms represented by dashed lines \cite{agv}.

\tbf Definition:
We call a variable $X$ a root if the polynomial $W$ contains a term
$X^a$.
A monomial $Y^aZ$ is called a pointer at $Z$.
$a$ is called the exponent of $X$ or $Y$, respectively.
We recursively define
a link between two expressions, which may themselves be variables or links,
as a monomial depending only on the variables occurring in these expressions.
A link may further be linear in an additional variable $Z$,
which does not count as a variable of the link. In this case we
say that the link points at $Z$, thus extending the previous definition
of a pointer. Of course a specific monomial occurring in $W$ can have
more than one interpretation as a link or pointer.
Given $W$, we call any graph (not necessarily the maximal one) whose lines
allow the above interpretation in terms of monomials in $W$ a graphic
representation of $W$.

\tbf Theorem 1:
For a configuration a necessary and sufficient condition for
non-degeneracy is that
it has a member which
can be represented by a graph with:\\
1. Each variable is either a root or points at another variable.\\
2. For any pair of variables and/or links pointing at the same variable Z
   there is a link joining the two pointers
   and not pointing at $Z$ or any of the targets of the sublinks
   which are joined.

Before proving this theorem, we shall illustrate the ideas on which it
is based with some examples.\\
{\tbf Example 1:} A polynomial of the form
\beq X_1^{\a_1}X_2+X_2^{\a_2}+X_2X_3^{\a_3}+X_4^{\a_4}X_5+X_5^{\a_5}+
X_5X_6^{\a_6}+X_7^{\a_7}+\e_1X_1^{p_1}X_3^{p_3}X_7+\e_2X_4^{p_4}X_6^{p_6}X_7
						       \eeq
is degenerate, as one can see by calculating $dW$ and setting $X_2,\,X_5$
and $X_7$ to 0. By adding the ``missing link''
$\e_3X_1^{q_1}X_3^{q_3}X_4^{q_4}X_6^{q_6}$ we obtain non-degeneracy. The
translation into our graphic language is given by {\it Fig. 4}. \putfig\figd\\
{\tbf Example 2:} {\it Fig. 5}                                  \putfig\fige
shows the graphic representation of a polynomial
where more than two arrows end at the same point. Distinct ``links for links''
are realised by the same monomial. \\
{\tbf Example 3:}
$W=X_1^3+(X_2^2+X_3^2+X_4^2)X_1+\e X_2X_3X_4$. Note that the last monomial
fulfils three tasks at once: pointer from $X_2$ and $X_3$ to $X_4$, pointer
from $X_2$ and $X_4$ to $X_3$ and pointer from $X_3$ and $X_4$ to $X_2$.

{\tbf Proof of Theorem 1:}
A) `Necessary': Calculating $dW/dX_i$ yields $n$ equations in
$n$ variables.  Non-degeneracy means that they can be fulfilled only by
the trivial solution. Obviously a necessary condition is that by setting
$k$ variables to $0$ no more than $k$ equations can be fulfilled
identically. Starting with $k=n-1$, we find that for each $X_i$ monomials
in this variable must occur in $dW/dX_j$, which means that
each variable has to be a root or has to point at another variable.
(This was the first condition of the theorem).\\
Let us now consider $k=n-2$, i.e. we set to 0 all variables except two
(which we shall call $X_1$ and $X_2$). According to above, monomials
$X_1^{\a_1}$
and $X_2^{\a_2}$ must occur. There are two possibilities: either these
monomials
occur in different equations, meaning that there are at
least two equations that are not automatically fulfilled by setting
all variables except
$X_1$ and $X_2$ to zero, or they both occur in the same equation.
In the latter case a necessary condition for non-degeneracy is the
occurrence of a monomial in both variables in one of the other equations,
i.e. either one of them points at the other or there must be a link
between them. \\
We proceed inductively in $l=n-k$: We construct the graph by adding
just
the lines that we need in each step. Assume we have all the lines up to level
$l-1$. Suppose there are two links pointing at the same variable $Z$.
Let them have  $l_1<l$ and $l_2<l$ variables, of which they have $q$ variables
in common, and $l=l_1+l_2-q$ variables together. Setting all other variables
to zero, the links and pointers we have drawn until now correspond to
$l_1+l_2-q-1=l-1$ equations, as $\6W/\6Z$ and $q$ additional equations
are double counted due to the overlap in the variables.
So we need one more
equation in the $l$ variables, i.e. we need the link implied by the second
condition of the criterion.\\
B) `Sufficient': We show that no degenerate configuration can fulfil our
conditions. Let $W=\sum M_\m\e^\m$ represent a degenerate configuration,
where the $M_\m$ are monomials in $X_i$, and let $\6_i=\6/\6X_i$ and
$\6_\m=\6/\6\e^\m$. Further, for any choice of $\e^\m$ there are non-vanishing
solutions $X_i(\e)$ to the equations $\6_iW=0$. As these equations are
polynomial, we can choose an open set of $(\e^1,\ldots,\e^k)$ and a solution
$(X_1(\e),\ldots,X_n(\e))$ depending smoothly on the $\e^\m$ in this set.
Quasihomogeneity implies that $W(X_i(\e))\eq0$. Therefore
\beq     {d\0d\e^\m}W=\sum_i{\6_i W}{\6_\m X_i}+{\6_\m W}=M_\m=0,         \eeq
i.e. every monomial in $W$ has to vanish. We  now choose a point $\e$ in this
set for which $X_1(\e)\neq0,\ldots,X_l(\e)\neq0$, but $X_{l+1}\eq\ldots\eq X_n
\eq0$ in some neighbourhood of $\e$ (we are of course free to choose the
labels for the $X_i$). Differentiating $\6_iW$ with respect to $\e^\m$ we find
\beq     {d\0d\e^\m}\6_iW=\sum_{j\le l} \6_i\6_jW \6_\m X_j+\6_iM_\m=0.   \eeq
Due to quasihomogeneity $\sum_{j\le l}q_jX_j\6_i\6_jW=(1-q_i)\6_iW=0$ , i.e.
the rank of the rectangular matrix $\6_i\6_jW$ with $j\le l$ is less than $l$.
Thus there are at least $n-l+1$ independent vectors $c_i^{(m)}$ with
$\sum_{i\le n}c_i^{(m)} \6_i\6_jW=0$ and hence $\sum_{i\le n}c_i^{(m)}\6_iM_\m
=0$. As all monomials $M_\m$ have to vanish, only one of the derivatives
$\6_iM_\m$ can be non-vanishing for a given $\m$. Thus the sum
$\sum_ic_i^{(m)}\6_iM_\m$
has at most one non-vanishing contribution, and, as $c_i^{(m)}$ can be 0 for
all $m$ for at most $l-1$ values of $i$, all variables $X_1,\ldots,X_l$
and all links among these variables have
to point at a subset of at most $l-1$ variables $X_j$ with $j>l$. The
resulting double pointer cannot be completely resolved, as the required link
would again have to point at the same set of $l-1$ variables and thereby
generate a new double pointer.
 //.

{\tbf Lemma 1:} The necessary condition for non-degeneracy that
the expression (\ref{pp}) for the Poincar\'e polynomial is a polynomial
(we refer to this as the PP-condition)
is equivalent to the criterion of theorem 1 if one omits the
requirement that all exponents in the link monomials have to be non-negative.

{\tbf Proof:} $\prod(1-t^{d-n_i})/(1-t^{n_i})$ is a polynomial if all zeros
in the denominator, counted according to their multiplicities,
are matched by zeros
in the numerator, i.e. if the set of all multiples of $1/n_i$ between 0
and 1 is a subset of the multiples of $1/(d-n_i)$, even when multiplicities
are taken into account. The relaxed condition on the links, which we are
referring to, is that the number theoretic condition $\sum p_in_i=d$
(for non-pointing links) or $\sum p_in_i=d-n_k$ (for a link pointing at
$X_k$), where the sum runs over all $X_i$ joined by the link, is fulfilled
without requiring that the $p_i$ all be non-negative. This is equivalent to
the condition that the greatest common divisor of these $n_i$ divides $d$ or
some $d-n_k$. To show the equivalence of these two conditions we first note
that the PP-condition implies that each $X_i$ has to be a pointer or a root,
as $1/n_i$ must be a multiple of some $1/(d-n_k)$. In this context a root is
to be considered as a pointer at itself. \\
There is a problem, however, if two variables $X_i$ and $X_j$ point at the
same variable $X_k$. Then the multiples of $1/(n_i\cap n_j)$ are contained
only once and have to be taken care of by some further variable $X_l$ with
$n_i\cap n_j$ dividing $d-n_l$. We thus recover the requirement of the links
implied by the non-degeneracy criterion except for the
positivity of the exponents in the monomial. If $l$ cannot be chosen as $i$
or $j$, i.e. if $n_i\cap n_j$ does not divide $d$, this link is a pointer at
$X_l$.
The roots, considered as pointers at themselves, and the links which are no
pointers, i.e. can be considered as pointing at one of their variables,
do not imply additional links. This is because the corresponding $n_i$
(or $n_i\cap n_j$ for pointers) are divisors of $d$. In the
simplest case, for example, where $n_i$ and $n_j$ divide $d-n_i$, the missing
ratio $1/(n_i\cap n_j)$ is a multiple of $1/(d-n_j)$ and thus acts like a
pointer at $X_j$ with the additional feature that the denominator divides $d$.
In this way the missing numbers can be passed on backwards along any pointers
with ever smaller denominators until they find their match in a free multiple
of some $1/(d-n_k)$ with no other pointer at
$X_k$ or until $n_i\cap\ldots\cap n_k=1$.
Proceeding in the same way with the additional overlaps which may arise
due to
the additional pointers in each step, we indeed find equivalence of the
PP-condition and the relaxed link criterion. //.

To illustrate this equivalence we use the following example by B. M. Ivlev
of a degenerate configuration fulfilling the PP-condition \cite{agv},
\beq  \IC_{(1,24,33,58)}[265],  \eeq
with the corresponding skeleton polynomial
\beq X^{265}+XY^{11}+XZ^{8}+ZU^4. \eeq
To fulfil the criterion of theorem 1 we would
need a link between $Y$ and $Z$, which has to point at $U$ as neither
$d$ nor any $d-n_i$ is a multiple of $3=24\cap33$ except for
$d-n_U=207=24p+33q$. This equation, however, does not have a solution with
both $p$ and $q$ positive, which explains why (\ref{pp}) is a polynomial
although the configuration is degenerate.
It would be interesting to find out if it is possible to construct
an N=2 SCFT with these weights.

\section{On configurations with a given index}

{\bf Lemma 2:} If a non-degenerate configuration contains $n$
variables $X_i$ with
a given weight $q\in(1/3,1/2)$, then it also contains at least $n$
variables $Y_j$ of weight $\bar q=1-2q$.

{\tbf Proof:} Let $W=W(X_i, Y_j, Z_k)$ with $weight(X_i)=q$, $weight(Y_j)=
\bar q$ and $weight(Z_k)\neq q,\bar q$. We calculate $dW$ and set $Y_j$
and $Z_k$ to 0. Non-degeneracy implies that at least $n$ equations for the
$X_i$ must remain. Because $q\in(1/3,1/2)$, these equations must be
quadratic in the $X_i$, i.e. of weight $2q=1-\bar q$. Therefore they must
come from $dW/dY_i$, implying that the number of $Y_i$'s is at least $n$. //.

We call variables trivial if they correspond to terms $X^2$. Trivial variables
have weights $q=1/2$ and therefore do not contribute to $\ch$,
nor to the local algebra, as they can be eliminated by $\6W/\6X=0$.

{\tbf Corollary 1:} For every non-degenerate configuration $\ch$ is
greater than or equal to 1/3 times the number of non-trivial variables.

{\tbf Proof:} Grouping variables with $q\in(1/3,1/2)$ together with $\bar q$,
we have $1-2q+1-2\bar q=2q>2/3$; the contributions of all other
non-trivial variables (with weights $\le1/3$) are at least $1-2/3=1/3$. //.

{\tbf Theorem 2:} Given a positive rational number $\ch$, there is only a
finite number of non-degenerate configurations whose index is $\ch$.

{\tbf Proof \hbox{\rm(indirect)}: Suppose there is an infinite sequence of
configurations with index $\ch$. Due to the above
corollary there is only a finite
number of skeleton graphs that can realize $\ch$, so there must be an infinite
subsequence corresponding to just one graph. Considering a specific exponent,
we can find either a subsequence for which this exponent is constant or one
for which this exponent goes monotonically to infinity. Doing this for every
exponent, we end up with a sequence of polynomials
corresponding to the same
skeleton graph, for which $n-m$ exponents are constant while $m$ exponents
tend monotonically to infinity. Of course the ``limit configuration'', which
contains $m$ variables of weight 0, also has index $\ch$. We denote the
weights in the $l$'th member of the sequence by $q_i^{(l)}$,
$i\in \{1,\cdots,n\}$ with
$\lim_{l\to\infty}q_i^{(l)}=0$ for $i\in\{1,\cdots,m\}$. We will now show that
the index of a member of such a sequence is in fact smaller than the index
of the limit sequence, in contradiction with the assumed constance of $\ch$.
Consider a specific $l$ and let $\e=\max
q_i^{(s)}$ with $i\le m$ and $s\ge l$. We define
the intervals $I_k=(2^{-k}\e, 2^{-k+1}\e]$. By $A_k$ we denote the number of
points with $q^{(l)}-q^{(\infty)}\in I_k$, by $B_k$ we denote the number of
points with $q^{(\infty)}-q^{(l)}\in I_k$. Note that $B_1=0$,
as for any variable
$q^{(\infty)}=0$ or the possible target variable has $q<\2$.
Now consider all $\sum_{i=1}^kB_i$
points for which $q^{(\infty)}-q^{(l)}>2^{-k}\e$. Setting all other variables
to zero in $dW/dX_i$, we see that we need at least as many equations, coming
from points with $q^{(l)}-q^{(\infty)}>2^{-k+1}\e$,
as these equations are at least quadratic in the non-zero variables.
We thus have $\sum_{i=1}^k B_i\le \sum_{i=1}^{k-1} A_i$,
i.e. $\sum_{i=1}^k (A_i-B_{i+1})\ge 0$ and
\beq \ch^{(\infty)}-\ch^{(l)}=2\sum(q^{(l)}-q^{(\infty)})>2\e\sum_i(2^{-i}A_i
    -2^{-i+1}B_i)=2\e\sum_i2^{-i}(A_i-B_{i+1})\ge 0.                  \eeq
This is the contradiction we were looking for. //.

}
{\tbf Theorem 3:} Given a rational number $r$,
one can find a positive number $\e$
such that no number in the interval $(r,r+\e)$ is the index of a
non-degenerate quasihomogeneous polynomial; i.e. the accumulation points
in this set of indices are all approached from below.

{\tbf Proof:} Like the previous theorem
with an infinite sequence of configurations
with decreasing indices $\ch^{(l)}\to r$ instead of constant index. //.

Theorem 2 has given us valuable theoretical information, but
does not help us in explicitly finding all configurations of a given index.
Especially for the case $\ch=3$, which is most important for string theory,
one would like to have a way of constructing normal forms for all
possible configurations. We will now
formulate and prove a lemma which makes it possible to write a computer
program which calculates explicitly all configurations with $\ch=3$.

{\tbf Lemma 3:} For a non-degenerate quasihomogeneous polynomial
with $\ch=3$ the number of exponents $\a_i>18$ or $\a_i>84$
is smaller than 3 or 2, respectively.
These limits cannot be improved, as is seen from the polynomials
$X_1^3+X_2^{18}+X_3^{18}+X_4^{18}$ and $X_1^3+X_2^7+X_3^{84}+X_4^{84}$.

{\tbf Proof:} Let us assume that there are three exponents $\a_i>18$,
corresponding to variables with weights
$q_1,q_2,q_3<1/18$, contributing $1-2q_i$ to $\ch$.
If there are also variables with exponents $\a=2$, pointing at variables with
$q_i<1/18$, they add $1-2(1/2(1-q_i))=q_i$ to $\ch$. The total contribution
of the variables considered so far is therefore between $3(1-2/18)=3-1/3$
and 3. It is impossible to complete this to a polynomial with
$\ch=3$. This proves the first assertion. \\
Assume now that there are terms $X_i^{a_i}$ or $X_i^{a_i}Y^i$
with $a_i>84$, $i=1,2$ in the potential, thus $q_i<1/84$.
The contribution $\ch_{12}$ of these variables to $\ch$ fulfils
$2-{1\021}<\ch_{12}=2-2q_1-2q_2<2$. A ``partner variable'' $Z_i$ pointing at
$X_i$
with exponent 2, i.e. a term $X_iZ_i^2$, can be disregarded as it effectively
just doubles $a_i$ (if $a_1=a_2$ both possible partner variables might point
to the same $X_i$). According to lemma 2 there have to be 2 or 3 more
variables in order to make up for the difference to a total $\ch$ of 3.\\
{\it Case 1:} 2 additional variables $U_i$ with weights $r_i$. If one of their
exponents is 2, it has to point at a $Z_i$, since $U_1$ and $U_2$ have to
contribute
more than 1. Then one can explicitly calculate the contribution of $X_1, X_2,
Z_1, (Z_2,) U_1$ to be ${5\0 2}-{3\0 2}q_1-(2)q_2$. $\ch=3$ would then
require $r_2={1\0 4}-{3\0 4}q_1-(\2)q_2$, which cannot be satisfied.
The exponents of $U_1,U_2$ also have to be less than 7, since
${1\0 3}+1-{2\0 7}={22\021}$. Enumeration of all relevant singularities in 2
variables
shows that the smallest possible contribution to $\ch$ larger than 1 is 22/21,
thus the total $\ch$ cannot be 3. If one or both $U_i$ point at some $X_i$ or
$Z_i$, $\ch$ can only be enlarged. So we finally have to consider the
contributions with $\ch=2-2(r_1+r_2)\le1$. Pointers at $X_j$ 
would not make any difference, as
the decrease in $r_i$ would be $r_iq_j$, which is not sufficient to reach
$\ch=3$. So let $U_1$ point at $Z_j$. This makes $1-2r_1$ larger than
${2\03}-{1\0252}$. Thus $U_2$ cannot point at a $Z_k$ and its exponent has to
be 3, so $U_2$ cannot make up for the difference to $\ch=3$.\\
{\it Case 2:} 3 additional variables $U_i$ with weights $r_i$. We split this
case according to the number of exponents equal to 2. If all exponents are
larger than 2 this implies $r_i\le{1\0 3}$, and hence $r_i\ge{13\0 42}=
{1\0 3}(1-{1\0 14})$, i.e. all
exponents have to be 3. These variables thus may only point at a variable with
weight less than $1/14$, i.e. at $X_i$. Examining all cases there are
10 infinite series of polynomials with $\ch=3$: \\
\def\seri#1#2{\hbox to 108mm{\hskip 20mm #1\hfil } and \hskip 3mm #2 \\}
\seri{$X^a+Y^{6a}+XZ^2+U^3X+V^3X+W^3$ }{ $Y^{6a} \to Y^{3a}+YT^2$}
\seri{$X^a+Y^{4a}+XZ^2+U^3X+V^3X+W^3Y$ }{ $Y^{4a} \to Y^{a}+YT^2$}
\seri{$X^a+Y^{2a}+XZ^2+U^3X+V^3X+W^3X$ }{ $Y^{2a} \to Y^{a}+YT^2$}
\seri{$X^a+Y^a+X(U^3+V^3+W^3+Z^2+T^2)$ }{ $XW^3 \to YW^3$}
\seri{$X^a+Y^a+X(U^3+V^3+W^3)+Y(Z^2+T^2)$ }{ $XW^3 \to YW^3$}
For none of these polynomials, however, do the links which would be
necessary to make them non-degenerate exist for $a>84$.
In the first case, for example, it is impossible to have a link between
$Z$ and $U$, since the weights of these two variables add up to ${5\0 6}
(1-{1\0 a})$. \\
Now let the first exponent be equal to 2. $U_1$ must not point
at $Z_i$ (otherwise $\ch>3$), so it points at $U_2$ and the exponents of
$U_2$ and $U_3$ have to be 3 as above. If the latter variables do not point
further the total $\ch$ stays below 3. They might only point at $X_1$, because
otherwise we would violate $\ch\le3$. This, however, does not help either,
because the decrease of $r_2$ and/or $r_3$ is sufficiently compensated by
the increase of $r_1$ to keep $\ch$ below 3. The proof is completed by the
observation that there cannot be 2 exponents of the $U$'s equal to 2, because
one of these variables would then have to point at a $Z_i$, which would
increase
$\ch$ beyond 3 for an isolated configuration. //.

Lemma 3, together with corollary 1,
implies an algorithm for the calculation of all configurations
with $\ch=3$, as the only unconstrained exponent in each skeleton graph
can be calculated (and of course has to turn out integer to yield a solution).
It is straightforward to further reduce the number of possibilities using
lemma 2. Consider, for example, a skeleton graph with $n$ points of which
$i$ have an exponent larger than or equal to $a$. If
$p$ of these $i$ variables
are pointed at by a variable with exponent 2, this implies
\beq\chat\ge(n-i-p)/3+i(1-2/a)+p/a. \eeq
If $n+2i>3\ch+p$ this is equivalent to
\beq a\le3(2i-p)/(n+2i-3\chat-p). \eeq
As $\6a/\6p$ is always positive, and obviously $p\le i$ and $p\le n-i$,
we obtain the bounds $a\le3i/(n+i-3\ch)$ or $a\le (3i-n)/(i-\ch)$ for
$i\le n/2$ or $i\ge n/2$, respectively. It is also straightforward to check
that not more than $i$ variables can have an exponent larger than
$a=2i/(i-\ch)$.

For completeness we state the following results referring to $\ch=1$ or 2.
They can be proved with the same methods as above.

{\tbf Lemma 4:} Any configuration with $\ch=1$ corresponds to weights
$(1/3,\,1/3,\,1/3)$, $(1/3,\,1/6)$ or $(1/4,\,1/4)$.
For a polynomial with $\ch=2$ there is
at most one exponent greater than 12 and no exponent can be greater than 42.

\old
\tbf Lemma: In a potential in 8 variables with $\ch=3$ no exponent is
larger than 15.\\
{\bf Proof} (indirect): Suppose there is such a variable $X$ with weight
$q$.\\
A) $X$ has no partner variable. All other variables fulfil 1/3-Lemma,
$\ch>3$.\\
B) $X$ has more than one partner, implying there are more variables of weight
 $q$.
The contribution of two such variables (plus partners) to $\ch$ is greater
than $2(1-q)>28/15$, implying $\ch>3$.\\
C) $X$ has one partner, another variable pointing at the partner. Worst case:
$XZ_1^2+Z_1Z_2^2+Z_2Z_3^2$, contributes $(7-5q)/4$ to $\ch$. For $q<1/15$
this is greater than 5/3, implying $\ch>3$. (That's the sharp one. For
all other purposes $\a\le 12$ is sufficient).\\
D) $X$ has one partner, no variable pointing at the partner. $X$ and its
partner contribute $1-q\in (1-1/15,1)$ to $\ch$.\\
a) $XZ_1^2+XZ_2^3$ allows no link between $Z_1$ and $Z_2$ unless there is
another variable of weight $q$ - see case B)\\
b) $XZ_1^2+XZ_2^\b$ with $\b\ge 4$: Worst case  $XZ_1^2+XZ_2^4+Z_2Z_3^2$
yields $\ch>3$.\\
c) No variable pointing at $X$. If all other variables have weight 1/3, we're
below $\ch=3$. The two smallest deviations occur in $Y_1^3+Y_1Y_2^3+Y_2Y_3^2$
and $Y_1^4+Y_1Y_2^2$. Again we get $\ch>3$. //. \endold

\section{Conclusion}

We have given a criterion for the non-degeneracy of a configuration
which requires the check of a recursive condition concerning the
existence of certain
monomials consistent with quasihomogeneity, and have introduced a convenient
graphic representation for these monomials. The PP-condition is
equivalent
to this condition except for positivity of the exponents.
We have also shown that for a given singularity index the number of
non-degenerate configurations is finite and that such a value cannot be
approximated from above by non-degenerate configurations.
Finally, applications of these results provide the necessary
ingredients for explicit calculations of all configurations
at least for low values of $\hat c$.

Our results in particular imply an algorithm for the calculation of all
configurations with $\ch=3$, which is straightforward to implement
due to the recursive structure of the condition of theorem 1.
According to lemma 2 only polynomials in up to $3\ch$
non-trivial variables have to be considered.
As the weights of all variables are already determined by the
skeleton graph, one only has to investigate a reasonable number of such
graphs, which can be constructed recursively.
The crucial point is that the combinations of exponents which have to be
checked can be restricted to a finite number even for these skeletons,
as has been done in lemma 3 for the particularly interesting case $\ch=3$.
Non-degeneracy can then be checked in a second step.

Such a construction would be an extension of the work by Candelas et al.
\cite{cls} on Calabi-Yau manifolds in weighted projective spaces \cite{wps}.
The connection between these manifolds and
the construction of $N=2$ SCFT from non-degenerate
quasihomogeneous functions has been discussed by Greene et al. in
{\it ref.} \cite{gvw}.
The formulae for the calculation of the Hodge numbers from the scaling
dimensions of the superfields have later been supplied by Vafa \cite{v} and
rederived by methods of algebraic geometry in {\it ref.} \cite{roan}.
The authors of {\it ref.} \cite{cls} have implemented 30
polynomials fulfilling the criterion of theorem 1 and
have constructed some 6000 Calabi-Yau manifolds in weighted $\IP_4$.
As the number of skeleton graphs already grows faster than $2^n$ for
polynomials in $n$ variables, a complete construction along
these lines, however, appears to be difficult for larger numbers of variables.
\ifsub \ve \fi

\ifsub  {\newpage \noindent \Large\bf Figures} \par \pagestyle{empty}
      \figa \figb \figc \figd \fige \fi
\end{document}